\documentstyle[aps,twocolumn,epsf]{revtex}

\begin{document}


\title{Correlations derived from Modern Nucleon-Nucleon Potentials}
 
\author{H.\ M\"{u}ther} 
\address{Institut f\"ur Theoretische Physik, Universit\"at T\"ubingen,
         D-72076 T\"ubingen, Germany}
\author{A.\ Polls}
\address{Departament d'Estructura i Costituents de la Mat\`eria,
         Universitat de Barcelona, E-08028 Barcelona, Spain}

\maketitle

\begin{abstract}

Various modern nucleon-nucleon (NN) potentials yield a very accurate fit to
the nucleon-nucleon scattering phase shifts. The differences between these
interactions in describing properties of nuclear matter are investigated.
Various contributions  to the total energy are evaluated employing the Hellmann -
Feynman theorem. Special attention is paid to the two-nucleon correlation
functions derived  from these interactions. Differences in the predictions
of the various interactions can be traced back to the inclusion of non-local
terms. 

\end{abstract}


\section{Introduction}
The microscopic theory of nuclear structure based on realistic nucleon-nucleon
(NN) interactions is a very demanding subject because it requires the
description of a strongly correlated many-fermion system. Attempts to determine
e.g.~the energy of nuclei from a realistic NN interaction by using the
mean-field or Hartree-Fock approximation fail badly: such attempts typically
yield unbound nuclei. The strong short-range and tensor components of a
realistic NN interaction induce correlations into the many-body wavefunction
of nuclear systems. Many attempts have been made to measure these correlations
in detail. As an example for such measurements we mention the exclusive
$(e,e'NN)$ reactions, which have been made possible using modern electron
accelerators\cite{adam,rosner}. The hope is that the detailed analysis of such
experiments yields information about the correlated wavefunction of the nucleon
pair absorbing the virtual photon. This could be a very valuable test for the
model of the NN interaction producing these correlations.

In recent years, a new generation of realistic NN potentials 
has been developed which produce very accurate fits of the proton-proton
and proton-neutron (pn) scattering phase shifts\cite{nim,v18,cdbonn}. Since
these fits are based on the same phase shift analysis by the Nijmegen
group\cite{nimpsa} and yield a value for the $\chi^2$/datum 
very close to one, these various
potentials could be called phase-shift equivalent NN interactions.

This means that the on-shell matrix elements for the transition matrix $T$ are
essentially identical. This, however, does not imply that the underlying
potentials nor the effective interaction between off-shell
nucleons moving inside a nucleus are the same. Indeed it has been demonstrated
that these phase shift equivalent potentials yield different results even 
for the deuteron. Of course all of them reproduce the same empirical binding
energy and other observables, because these are part of the observables to which
the interaction has been fitted. However, the various contributions to the total
energy, the kinetic energy and the potential energy in the $^3S_1$ and $^3D_1$
partial waves, are quite different indicating that also the two-body
wavefunctions must be different\cite{deuter,deut1}. 

It is one of the aims of this study to explore whether similar differences 
can also be observed in calculating the energy of nuclear matter. These energies
are calculated using the Brueckner-Hartree-Fock (BHF) approximation. The BHF
approach for nuclear matter assumes a model-wavefunction of a free Fermi gas,
occupying plane wave states up to the Fermi momentum. The effects of
correlations are taken into account in terms of the Brueckner $G$-matrix.  
The BHF approximation does not give direct access to quantities like the kinetic 
or the potential energy. However, in the next section we will illustrate how  
the Hellman-Feynman Theorem \cite{helfr} can be used to  calculate these 
quantities and also the expectation value of the $\pi$ exchange
calculated for the correlated
wave function. The results for various 
contributions to the binding energy as well as the wave functions of 
correlated NN pairs in nuclear matter will be presented in section 3. The last
section contains a summary and conclusions.

\section{Correlations in Brueckner Hartree Fock} 

The central equation of the BHF approximation is the Bethe-Goldstone equation,
which defines an effective interaction $G$ for two nucleons in nuclear matter
occupying the plane wave states $i$ and $j$ by
\begin{equation}
G \vert ij >  =  V \vert ij > + V \frac{Q}{\epsilon_i + \epsilon_j - H_0} G 
\vert ij >\, .\label{eq:bethegol}
\end{equation}
Here and in the following $V$ stands for the bare NN interaction, $Q$ denotes
the Pauli operator, which prevents of the interacting nucleons into intermediate
states with momenta below the Fermi momentum $k_F$, $H_0$ defines the spectrum
of intermediate two-particle states and the BHF single-particle energies are
defined by
\begin{equation}
\epsilon_i = \frac{\hbar^2k_i^2}{2m} + \int_0^{k_F} d^3k_j <ij \vert G 
\vert ij >\,, \label{eq:epsbhf}
\end{equation}
as the sum of the kinetic energy of a free nucleon with mass $m$ and momentum 
$k_i$ and the potential energy. The single particle potential corresponds to the
Hartree-Fock approximation but calculated in terms of the effective interaction
$G$ rather than the bare interaction $V$. Also the total energy of the system is
calculated in a similar way containing the kinetic energy per nucleon of a 
free Fermi gas
\begin{equation}
\frac{T_{FG}}{A} = \frac{3}{5} \frac{\hbar^2 k_F^2}{2m}  \,,\label{eq:tfg}
\end{equation}
and the potential energy calculated in the Hartree-Fock approximation replacing
$V$ by the effective interaction $G$ (For a more detailed description see
e.g.~\cite{haftab}). This means that the BHF approach considers a model wave
function, which is just the uncorrelated wave function of a free Fermi gas and
all information about correlations are hidden in the effective interaction $G$.
Since this effective interaction is  constructed such that $G$ applied to the
uncorrelated two-body wave function yields the same result as the bare
interaction $V$ acting on the correlated wave function 
\begin{equation}
G \vert ij >  = V \vert ij >_{\mbox{corr.}}\,, \label{eq:vpsig}
\end{equation}
the comparison of this equation with (\ref{eq:bethegol}) allows the definition
of the correlated two-nucleon wave function as
\begin{equation}
\vert ij >_{\mbox{corr.}} = \vert ij > + \frac{Q}{\epsilon_i + \epsilon_j - 
H_0} G \vert ij > \,. \label{eq:psicor}
\end{equation}
This representation demonstrates that the correlated wave function contains the
uncorrelated one plus the so-called defect function, which in this approach
should drop to zero
for relative distances between the two nucleons, which  are larger than the
healing distance.

The BHF approach yields the total energy of the system including effects of
correlations. Since, however, it does not provide the correlated many-body 
wave function, one does not obtain any information about e.g.~the expectation
value for the kinetic energy using this correlated many-body state. To obtain
such information one can use the Hellmann-Feynman theorem, which may be
formulated as follows: Assume that one splits the total Hamiltonian into
\begin{equation}
H = H_0 + \Delta V 
\end{equation}
and defines a Hamiltonian depending on a parameter $\lambda$ by
\begin{equation}
H(\lambda ) = H_0 + \lambda \Delta V\,. 
\end{equation}
If $E_\lambda$ defines the eigenvalue of
\begin{equation}
H(\lambda ) \vert \Psi_\lambda > = E_\lambda\vert \Psi_\lambda >
\end{equation}
the expectation value of $\Delta V$ calculated for the eigenstates of the
original Hamiltonian $H=H(1)$ is given as
\begin{equation}
<\Psi \vert \Delta V \vert \Psi > = \left. \frac{\partial E_\lambda}{\partial
\lambda} \right|_{\lambda=1}\, . \label{eq:helfey}
\end{equation}
The BHF approximation can be used to evaluate the energies $E_\lambda$, which
also leads to the expectation value $<\Psi \vert \Delta V \vert \Psi >$
employing this eq.(\ref{eq:helfey}). In the present work we are going to apply
the Hellmann-Feynman theorem to determine the expectation value of the kinetic
energy and of the one-pion-exchange term $\Delta V = V_\pi$ contained in the
different interactions.

\section{Results and discussion}

The main aim of the work presented here is to investigate differences in nuclear
structure calculations originating from four different realistic NN
interactions, which are phase-shift equivalent. These four interactions are the
so-called charge-dependent Bonn potential (CDBonn)\cite{cdbonn}, the Argonne
V18 (ArV18)\cite{v18} and the versions I (Nijm1) and II (Nijm2) of the Nijmegen
interaction\cite{nim}. All these models for the NN interaction  include a
one-pion exchange (OPE) term, using essentially the same $\pi NN$ coupling
constant, and account for the difference between the masses of the charged
($\pi_\pm$) and neutral ($\pi_0$) pion. However, even this long range part of
the NN interaction, which is believed to be well understood, is treated quite
differently in these models. The Nijmegen and the Argonne V18 potentials use
the local approximation, while the pion contribution to the CDBonn potential is
derived in a relativistic framework assuming pseudoscalar coupling. It has
recently been shown that the non-localities included in the relativistic
description of the CDBonn potential tends to lead to smaller D-state
probabilities in the deuteron\cite{deut1}.  

The description of the short-range part is also different in these models. The
NN potential Nijm2 \cite{nim} is a  purely local potential in the sense that it
uses the local form of the OPE potential for the long-range part and
parameterizes the contributions of medium and short-range distances
in terms of local
functions (depending only on the relative displacement between the two
interacting nucleons) multiplied by a set of spin-isospin operators. The same
is true for the Argonne $V_{18}$ potential \cite{v18}. The NN potential denoted
by Nijm1 \cite{nim} uses also the local form of OPE but includes a $\bf p^2$
term in the medium- and short-range central-force (see Eq.\ (13) of Ref.\
\cite{nim}) which may be interpreted as a non-local contribution to the central
force. The CD-Bonn is derived in the framework of the relativistic meson field
theory. It is calculated in momentum space and contains non-local  terms in the
short-range as well as long-range part including the pion-exchange
contribution.

First differences in the prediction of nuclear properties obtained from these
interactions are displayed in table~\ref{tab1} which contains various
expectation values calculated for nuclear matter at the empirical saturation
density, which corresponds to a Fermi momentum $k_F$ of 1.36 fm$^{-1}$. The most
striking indication for the importance of nuclear correlations beyond the mean
field approximation may be obtained from the comparison of the energy per
nucleon calculated in the mean-field or Hartree-Fock (HF) approximation. All
energies per nucleon calculated in the (HF) approximation are positive.
therefore far away from the empirical value of -16 MeV. Only after inclusion of
NN correlations in the BHF approximation results are obtained which are close to
the experiment. While the HF energies range from 4.6 MeV in the case of CDBonn 
to 36.9 MeV for Nijm2, rather similar results are obtained in the BHF
approximations. This demonstrates that the effect of correlations is quite
different for the different interactions considered. However it is worth noting
that all these modern interactions are much ``softer'' than e.g.~the old Reid
soft-core potential\cite{reid} in the sense that the HF result obtained for the
Reid potential (176 MeV) is much more repulsive.

Another measure for the correlations is the enhancement of the kinetic energy 
calculated for the correlated wave function as compared to the mean field
result which is identical to $T_{FG}$, the energy per particle of the free Fermi
gas. At the empirical density this value for $T_{FG}$ is 23 MeV per nucleon.
One finds that correlations yield an enhancement for this by a factor which
ranges from 1.57 in the case of CDBonn to 2.09 for Nijm1. It is remarkable that
the effects of correlations, measured in terms of the enhancement of the kinetic
energy or looking at the difference between the HF and BHF energies, are
significantly smaller for the interactions CDBonn and Nijm1, which contain
non-local terms.  

The table~\ref{tab1} also lists the expectation value for the pion-exchange
contribution $V_\pi$ to the two-body interaction. Here one should note that the
expectation value of $V_\pi$ calculated in the HF approximation is about 15 MeV
almost independent of the interaction considered. So it is repulsive and
completely due to the Fock exchange term. If, however, the expectation value for
$V_\pi$ is evaluated for the correlated wave function, one obtains rather
attractive contributions ranging from -22.30 MeV per nucleon (CDBonn) to -40.35
MeV (ArV18). This expectation value is correlated to the strength
of the tensor force or the D-state probability $P_D$ calculated for the
deuteron (see table~\ref{tab1} as well). Interactions with larger $P_D$, like 
the $ArV18$, yield larger values for $<V_{\pi}>$. For a further support of this
argument we also give the results for three different version of
charge-independent Bonn potentials A, B and C, defined in \cite{rup}. 

All this demonstrates that pionic and tensor correlations are very
important to describe the binding properties of nuclei. In fact, the gain in
binding energy due to correlations from $V_\pi$ alone is almost
sufficient to explain the difference between the HF and BHF energies. 

\begin{figure}[t]
\epsfysize=12.0cm
\begin{center}
\makebox[16.4cm][c]{\epsfbox{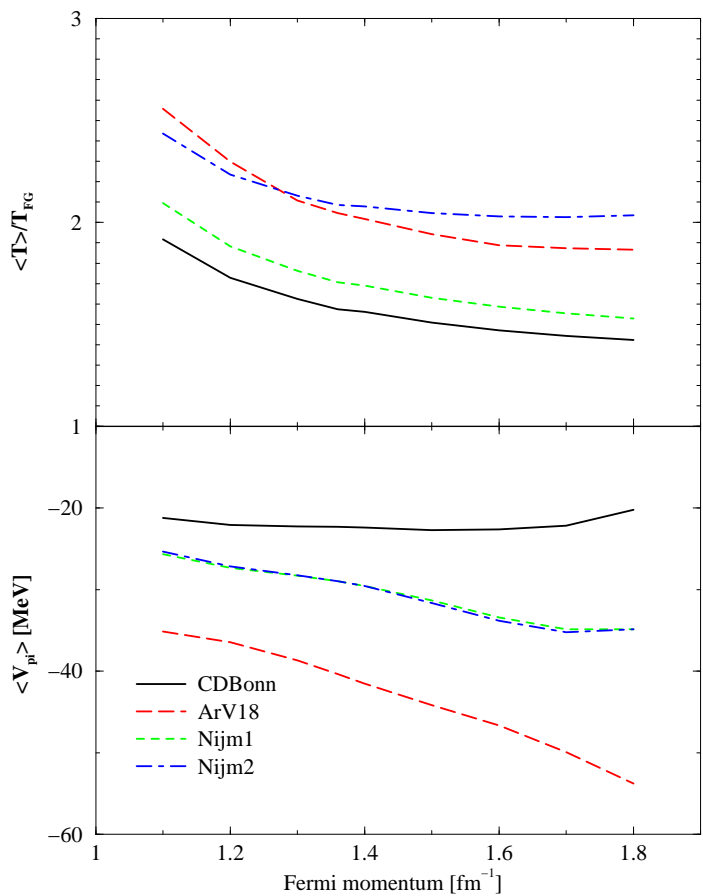}}
\end{center}
\caption{The upper part of this figure displays the ratio of the kinetic energy 
per nucleon, calculated for the correlated state, to the energy per nucleon of a 
free Fermi-gas (\protect\ref{eq:tfg}) as a function of the Fermi momentum. The 
lower part exhibits the expectation value of the one-pion-exchange contribution 
to the binding energy per nucleon. Different realistic NN interactions are 
considered. \label{fig1}}
\end{figure}

Until now we have just discussed results for nuclear matter at one density.
The values for the kinetic energy, $<T>$, and $<V_{\pi}>$ are displayed for
various densities in Fig.~\ref{fig1}. One finds that the ratio of the kinetic 
energy calculated for the correlated wave function, $<T>$, and the energy of 
the free Fermi-gas $<T_{FG}>$ decreases as a function of density. This plot
furthermore shows that the results for the different interactions can be 
separated in two groups: the local interactions, ArV18 and Nijm2, yield larger 
kinetic energies than CDBonn and Nijm1, which contain nonlocal terms.

The lower part of Fig.~\ref{fig1} shows that the pionic contribution to the 
total energy is quite different for the interactions. It is strongest for ArV18, 
getting more attractive for larger densities. The pionic contribution obtained 
from the other potentials is weaker and does not exhibit this increase at high 
densities. This may indicate that the enhancement of pionic correlations, which 
has been discussed in the literature as an indication for pion 
condensation\cite{pand}, is a feature which may not be reproduced by realistic 
interactions different from the Argonne potentials.

\begin{figure}[h] 
\epsfysize=9.0cm
\begin{center}
\makebox[16.4cm][c]{\epsfbox{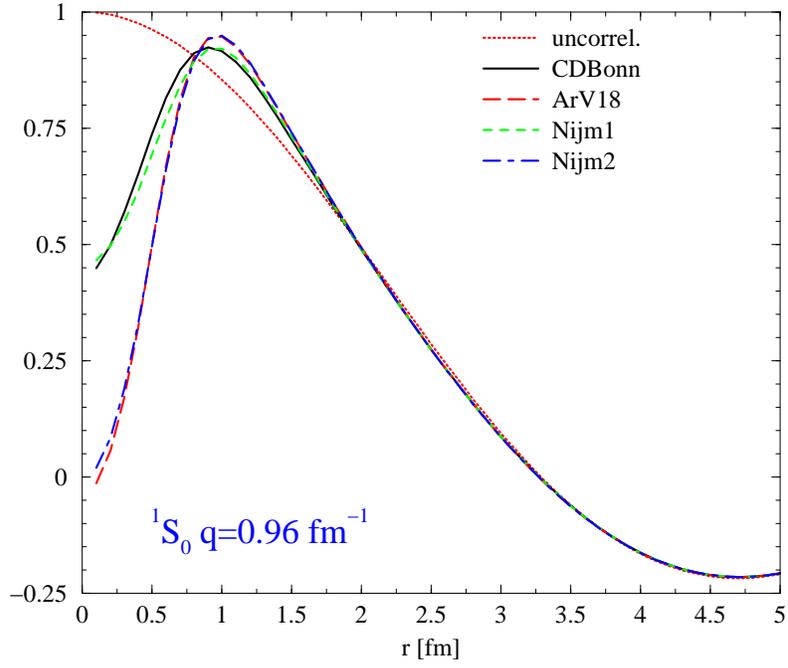}}
\end{center}
\caption{Correlated wave functions  $\vert ij >_{\mbox{corr.}}$ as defined in 
(\protect\ref{eq:psicor}) as a function of the relative distance for the $^1S_0$ 
partial wave. Results are shown for a pair of nucleons in nuclear matter at 
empirical saturation density, which heal to an uncorrelated two-nucleon wave 
function with momentum $q$ = 0.96 fm${-1}$ at larger distances. The curves are 
labeled by the interactions, which were considered. \label{fig2} }
\end{figure}

\begin{figure}[h] 
\epsfysize=9.0cm
\begin{center}
\makebox[16.4cm][c]{\epsfbox{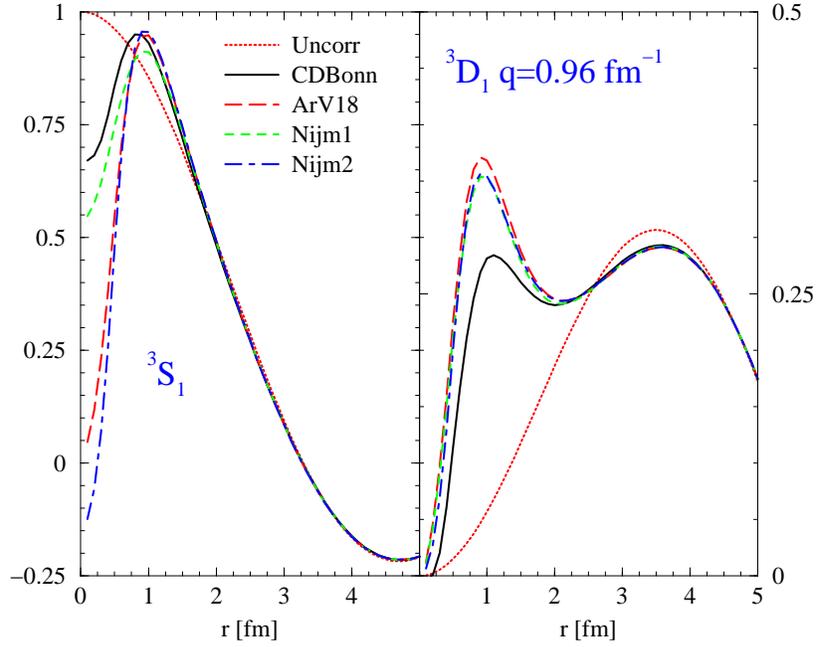}}
\end{center}
\caption{Correlated wave functions as a function of the relative distance for 
the $^3S_1$ and $^3D_1$ partial waves. Further details see 
Fig.~\protect\ref{fig2}. \label{fig3} }
\end{figure}

A different point of view on nuclear correlations may be obtained from 
inspecting the the relative wave functions for a correlated pair $\vert ij 
>_{\mbox{corr.}}$ defined in (\ref{eq:psicor}). Results for such correlated wave 
functions for a pair of nucleons in nuclear matter at empirical saturation 
density are displayed in Figs~\ref{fig2} and \ref{fig3}. As an example we 
consider wave functions which ``heal'' at larger relative distances to an 
uncorrelated two-nucleon wave function with momentum $q$ = 0.96 fm$^{-1}$ 
calculated at a corresponding average value for the starting energy.      
  
Fig.~\ref{fig2} shows relative wave functions for the partial wave $^1S_0$. One 
observes the typical features: a reduction of the amplitude as compared to the 
uncorrelated wave function for relative smaller than 0.5 fm, reflecting the 
repulsive core of the NN interaction, an enhancement for distances between 
$\approx$ 0.7 fm and 1.7 fm, which is due to the attractive components at medium 
range, and the healing to the uncorrelated wave function at large $r$. One finds 
that the reduction at short short distances is much weaker for the interactions 
CDBonn and Nijm1 than for the other two. This is in agreement with the 
discussion of the kinetic energies (see Fig.~\ref{fig1}) and the difference 
between HF and BHF energies (see table~\ref{tab1}). The nonlocal interactions 
CDBonn and Nijm1 are able to fit the NN scattering phase shifts with a softer 
central core than the local interactions.

Very similar features are also observed in the $^3S_1$ partial wave displayed in 
the left half of Fig.~\ref{fig3}. For the $^3D_1$ partial wave, shown in the
right part of Fig.~\ref{fig3}, one observes a different behavior: All NN
interactions yield an enhancement of the correlated wave function at $r\ 
\approx$ 1 fm. This enhancement is due to the tensor correlations, which couples
the partial waves $^3S_1$ and $^3D_1$. This enhancement is stronger for the
interactions ArV18, Nijm1 and Nijm2 than for the CDBonn potential. Note that the
former potential contain a pure nonrelativistic, local one-pion-exchange term,
while the CDBonn contains a relativistic, nonlocal pion-exchange contribution.

\begin{figure}[h] 
\epsfysize=9.0cm
\begin{center}
\makebox[16.4cm][c]{\epsfbox{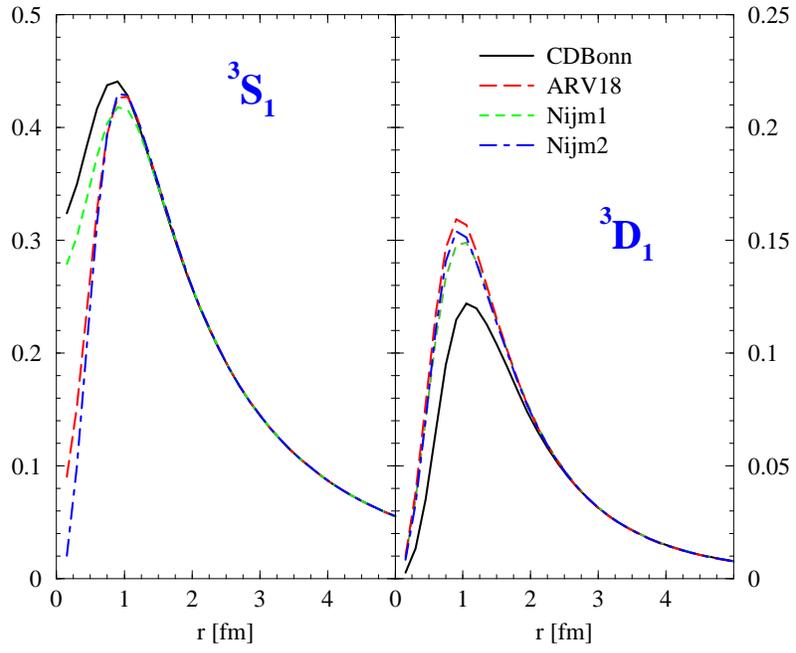}}
\end{center}
\caption{Wave function for the deuteron ($^3S_1$ and $^3D_1$) calculated for 
different realistic interactions. \label{fig4} }
\end{figure}

This behavior in the coupled  $^3S_1$ and $^3D_1$ waves can also be observed in
the corresponding wave functions for the deuteron, plotted in Fig.~\ref{fig4}.

\section{Conclusions}

Four modern NN interactions, the charge-dependent Bonn potential (CDBonn), the 
Argonne V18 (ArV18) and two versions of the Nijmegen potential (Nijm1 and 
Nijm2), which all give an excellent fit to NN scattering phase shifts, exhibit 
significant differences in calculating NN correlation functions and other 
observables in nuclear matter. Two of these interactions, CDBonn and Nijm1,  
contain nonlocal terms. These two interactions are considerably softer than the 
other interactions. This conclusion can be derived from three different 
observations: The Hartree-Fock energies are less repulsive, the kinetic energies 
calculated with the correlated wave functions are smaller and the correlated 
wave function in relative $S$ states are less suppressed at small relative 
distances.

The interactions also differ quite significantly in the pionic – or tensor 
correlations they induce. This is indicated to some extent by the deuteron wave 
function, in particular by the D-state probability. These differences, however, 
are  even enhanced in the nuclear wave functions leading to drastic differences
in the pionic contribution to the nuclear binding energy. The Argonne potential 
in particular yields a large pionic contribution, which increases with density. 
This importance of the pionic correlations is not observed for the other 
interactions.

It would be of great interest to study whether the differences between the 
correlations predicted from these interactions can be observed in experiments 
like the exclusive ($e,e`NN$) reactions in order to discriminate the various
models for the NN interaction.   

This work was supported in part by the SFB 382 of the Deutsche 
Forschungsgemeinschaft, the DGICYT (Spain) Grant PB95-1249 and the 
Program SGR98-11 from Generalitat de Catalunya.

\begin{table}
\begin{onecolumn}
\begin{tabular}{c|rrrr|rrrr}
& CDBonn & ArV18 & Nijm1 & Nijm2 & A & B & C & Reid\\
\hline
$<E>$ & -17.11 & -15.85 & -15.82 &  -13.93 & -16.32 & -15.32 & -14.40 
& -12.47\\
$<V>$ & -53.34 & -62.92 & -55.08 & -61.94 & -52.44 & -53.03 & -54.95 &
 -61.51\\
$<T>$ & 36.23 & 47.07 & 39.26 & 48.01 & 36.12 & 37.71 & 40.55 &
 49.04 \\
$<V_{\pi}>$ & -22.30 & -40.35 & -28.98 & -28.97 & -12.48 & -26.87 & -45.74 &
 -27.37 \\
$<E>_{\mbox{HF}}$ & 4.64 & 30.34 & 12.08 & 36.871  & 7.02 & 10.07 & 29.56 &
 176.25\\
\hline
$P_D$ [\% ] & 4.83 & 5.78 & 5.66 & 5.64 & 4.38 & 4.99 & 5.62 & 6.47\\
\end{tabular}
\caption{Energies calculated for nuclear matter with Fermi momentum $k_F$ = 1.36
fm$^{-1}$. Results are listed for the energy per nucleon calculated in BHF
($<E>$) and Hartree-Fock ($<E>_{HF}$) approximation.  Furthermore the
expectation value for the NN interaction $<V>$, the kinetic energy 
$<T_{\mbox{Kin}}>$ and the one-pion-exchange term $<V_{\pi}>$ are listed. For
completeness we also give the D-state probability calculated for the deuteron
$P_D$. Results are presented for the charge-dependent Bonn  (CDBonn)
\protect\cite{cdbonn}, the Argonne V18 (ArV18) \protect\cite{v18} and two
Nijmegen (Nijm1, Nijm2) \protect\cite{nim} interactions. For a comparison
results are also given for three older versions of the Bonn interaction (A,B,C)
\protect\cite{rup} and the Reid soft core potential \protect\cite{reid},
which is supplemented in partial waves in which it is not defined by the OBE C
potential. All energies are given in MeV per nucleon. \label{tab1}}
\end{onecolumn}
\end{table}

\vfil\eject

\vfil\eject

\end{document}